\documentclass[12pt]{article}
\usepackage{latexsym}
\usepackage{amsmath}
\usepackage{amsfonts}
\usepackage{amssymb}
\usepackage{amscd}
\usepackage{bbm}
\usepackage{fancybox}
\usepackage{cite}
\usepackage{amsmath,amsfonts,amsbsy}
\usepackage{pstricks,pst-node}
\usepackage[small,bf,hang]{caption2}
\usepackage{graphicx}
\usepackage{epsfig}
\usepackage{psfrag}
\usepackage{comment}

\usepackage{float}

\psset{unit=1.3cm,linewidth=.5pt,radius=.2}  

\usepackage{multirow}                     
\usepackage{float}                          
\usepackage{lscape}                         
\usepackage{bm}

\newcommand{\cO}{{\cal O}}

\newcommand{\beq}{\begin{equation}}
\newcommand{\eeq}{\end{equation}}
\newcommand{\bi}{\begin{itemize}}
\newcommand{\ei}{\end{itemize}}
\newcommand{\bt}{\begin{tabular}}
\newcommand{\et}{\end{tabular}}
\newcommand{\bc}{\begin{center}}
\newcommand{\ec}{\end{center}}

\newcommand{\vev}[1]{\langle#1\rangle}

\newcommand{\be}{\begin{equation}}
\newcommand{\ee}{\end{equation}}
\newcommand{\bea}{\begin{eqnarray}}
\newcommand{\eea}{\end{eqnarray}}
\newcommand{\ba}{\begin{array}}
\newcommand{\ea}{\end{array}}

\def\bbox{{\,\lower0.9pt\vbox{\hrule \hbox{\vrule height 0.2 cm
\hskip 0.2 cm \vrule height 0.2 cm}\hrule}\,}}
\newcommand{\dsl}{\pa \kern-0.5em /}

\begin{document}

\begin{titlepage}
\begin{center}

\hfill UG-11-56

\vskip 1.5cm

{\large \bf A Non-relativistic Logarithmic Conformal Field Theory \\ [.3truecm]
from a Holographic Point of View}

\vskip 1cm

{\bf Eric A.~Bergshoeff$^1$, Sjoerd de Haan$^1$, Wout Merbis$^1$, Jan Rosseel$^1$}

\vskip 25pt

{\em $^1$ \hskip -.1truecm Centre for Theoretical Physics,
University of Groningen, \\ Nijenborgh 4, 9747 AG Groningen, The
Netherlands \vskip 5pt }

{email: {\tt E.A.Bergshoeff@rug.nl, s.de.haan@rug.nl, w.merbis@rug.nl, j.rosseel@rug.nl}} \\
\end{center}

\vskip 1.5cm

\begin{center} {\bf ABSTRACT}\\[3ex]
\end{center}
We study a fourth-order derivative scalar field configuration in a fixed Lifshitz background. Using an auxiliary field we rewrite
the equations of motion  as two coupled second order equations. We specialize to the limit that the mass of the scalar field degenerates with
that of the auxiliary field and show that logarithmic modes appear.
Using non-relativistic holographic methods we calculate the two-point correlation functions of the boundary operators in this limit and find evidence for a non-relativistic logarithmic conformal field theory at the boundary.

\end{titlepage}

\newpage
%

\section{Introduction}

Gauge/gravity dualities have taught us a lot about the properties of strongly coupled field theories. The most studied gauge/gravity duality
is the AdS/CFT correspondence which  deals with the gravitational description of {\sl conformal} field theories \cite{Maldacena:1997re}.
Over the years generalizations of this correspondence have been proposed that are more closely connected to `real life' physical systems, like the quark-gluon plasma or, more recently, condensed matter systems.

 In the case of condensed matter physics, one conjectures a gravitational dual for  field theories that exhibit an anisotropic scale invariance. Such an anisotropic scaling behaviour can be embedded in various symmetry algebras, such as the Lifshitz algebra \cite{Kachru:2008yh}, which consists of spatial rotations and translations, time translations and a scaling transformation, and the Schr\"odinger algebra \cite{Son:2008ye, Balasubramanian:2008dm} which extends the former algebra with Galilean boosts and a number operator. Both algebras are characterized by a dynamical exponent $z$, which specifies how the scale transformations act differently on the time and spatial coordinates. Since algebras of this kind describe symmetries of non-relativistic field theories exhibiting non-relativistic dispersion relations, the corresponding  gauge/gravity dualities are often stated as giving examples of non-relativistic holography.

An independent new development in gauge/gravity duality is the connection between logarithmic conformal field theories (LCFT) and critical gravity theories \cite{Grumiller:2008qz,Skenderis:2009kd,Grumiller:2009mw,Grumiller:2009sn,Alishahiha:2010bw,Lu:2011zk,Deser:2011xc, Alishahiha:2011yb,Bergshoeff:2011ri,Porrati:2011ku,Lu:2011ks}. This connection was first made in the context of three-dimensional massive gravity theories, like Topologically Massive Gravity \cite{Deser:1981wh} or New Massive Gravity \cite{Bergshoeff:2009hq}. These are higher-derivative
 three-dimensional gravity theories, where the  Einstein-Hilbert action is supplemented with a negative cosmological constant and specific interactions with up to four derivatives. The spectrum of linearized perturbations of these theories is described by (unitary
 or non-unitary) massive bulk gravitons and boundary gravitons that do not describe any physical bulk degrees of freedom. At certain points in the parameter space of these theories, a degeneracy takes place and the massive gravitons coincide with the boundary gravitons. Such a special point  is dubbed a ``critical point'' and the theory at such a critical point is referred to as a ``critical gravity'' theory. At the critical point, the massive gravitons are replaced by so called logarithmic modes. According to the AdS/CFT dictionary, the boundary gravitons are dual to the components of the stress-energy tensor of the boundary field theory. The logarithmic modes on the other hand  source so-called logarithmic operators that degenerate with the components of the stress energy tensor in all quantum numbers. This results in a logarithmic conformal field theory, introduced in physics by \cite{Gurarie:1993xq}.\,\footnote{See \cite{Flohr:2001zs, Gaberdiel:2001tr} for reviews on LCFT and further references.} A defining feature of a LCFT is that the Hamiltonian is no longer diagonalisable: the components of the stress energy tensor form a pair with the logarithmic operators and the action of the Hamiltonian on such a pair is not diagonalisable.

Although the connection between LCFT's and critical gravity was discovered in three dimensions, it was found to hold also in higher-dimensional
higher-derivative gravity theories \cite{Lu:2011zk,Deser:2011xc, Alishahiha:2011yb,Bergshoeff:2011ri,Porrati:2011ku}. The mechanism by which logarithmic modes appear in the theory is similar to the three-dimensional case. At the critical point, a degeneracy takes place and the massive gravitons coincide with either massless gravitons or pure gauge modes. Instead of the massive gravitons, an equal amount of logarithmic modes appears in the theory.

In this letter, we wish to combine the two recent developments described above, i.e.~non-relativistic gauge-gravity duality and critical
gravity. To be precise, we propose a LCFT which enjoys anisotropic scale invariance. The approach we take in defining this `non-relativistic' LCFT is through the gauge/gravity duality: the LCFT is defined by its correlation functions, which we calculate through holographic methods
starting from a higher-derivative bulk theory.

Instead of looking on the gravitational side at a higher-derivative model of gravitational, i.e.~spin-2, degrees of freedom, we will consider a simpler situation involving only spin-0 degrees of freedom in a fixed non-relativistic gravitational background. This model is
a non-relativistic version of the model discussed in \cite{Ghezelbash:1998rj,Kogan:1999bn} (see \cite{Myung:1999nd} for a finite temperature version of this model). The model consists of a scalar field configuration in
a fixed AdS background with fourth-order derivative equations of motion and is conjectured to be dual to a LCFT. The higher-derivative equation
of motion can be written in terms of two coupled second order equations, involving Klein-Gordon operators, by introducing an auxiliary scalar field. At the point where the mass of the two scalar fields degenerate, logarithmic modes will appear. The boundary value of this logarithmic solution sources the logarithmic partner of the dual scalar operator and  defines a logarithmic conformal field theory on the boundary.
The analogy with critical gravity is as follows. A priori, the two Klein-Gordon operators involve different masses and the spectrum is described by two spin-0 excitations with different masses. At the critical point, both mass parameters are equal. Just as in critical gravity, massive gravitons coincide with massless gravitons and logarithmic modes appear, here both spin-0 excitations will coincide and a new logarithmic spin-0 mode shows up, that obeys the full fourth order equation of motion, but that is not annihilated by a single Klein-Gordon operator.

Since we are interested in a non-relativistic version of the model, we consider the background spacetime to be Lifshitz instead of AdS:
\begin{equation}\label{Lifshitz}
ds^2_{\rm Lif_{d+1}} = L^2\left( \frac{1}{r^{2z}}dt^2 + \frac{1}{r^2}dr^2 + \frac{1}{r^2} dx^a dx_a \right)\,.
\end{equation}
Here the $r$ and $x^a, a = 1, \cdots , d-1,$  are the spatial directions, $L$ is a parameter with inverse mass dimension and $z$ is the dynamical exponent. For $z=1$ we recover the relativistic AdS background. One can show that the Lifshitz spacetime has an anisotropic conformal boundary at infinity which can be mapped to $r=0$ \cite{Horava:2008ih,Horava:2009vy}. The bulk metric  induces an anisotropic conformal class of metrics on the boundary, where the action of the Lifshitz symmetry group on the boundary is induced from the action of the bulk isometries. 
The presence of logarithmic terms in representations of the Galilean Conformal Algebra and the Schr\"odinger-Virasoro algebra has been discussed in \cite{Hosseiny:2010sj,Hosseiny:2011ct}.

This paper is organized as follows. In  section \ref{model} we introduce the non-relativistic version of the model
 mentioned above and discuss some of its basic features. Furthermore, we give, for $z=2$, the logarithmic modes which source the logarithmic partner of the dual scalar operator. Next, in section \ref{twopoint} we derive the main result of this work. We use holographic renormalization to obtain the two-point functions of the dual operators in our non-relativistic model and indicate that they satisfy the defining properties of a non-relativistic LCFT. Finally, in the conclusions we discuss a few open issues and generalizations of our work.

\section{The Model} \label{model}

In this section we will introduce the scalar model that shares many of the features of critical gravity theories. It is, however,  much simpler
to study since it deals with spin-0 instead of spin-2 degrees of freedom. In subsection 2.1 we discuss some general features of the model
while in subsection 2.2 we will calculate the scalar logarithmic modes for the specific case $z=2$.

\subsection{General Features} \label{genfeat}

The model under consideration  consists of a scalar field $\phi_1$ obeying a fourth order equation of motion, given by the action of two Klein-Gordon operators on the field:
\begin{equation} \label{fourthorder}
\left( \Box - m_1^2 \right) \left( \Box - m_2^2 \right) \phi_1 = 0 \,.
\end{equation}
For $m_1^2 \neq m_2^2$, the solution space of this equation is spanned by the solutions of the two second order equations, obtained by acting with only one of the two Klein-Gordon operators appearing in (\ref{fourthorder}), i.e.~the full solution space is spanned by spin-0 excitations with masses $m_1$ and $m_2$. The case where $m_1^2 = m_2^2 = m^2$ is the analog of the critical point in massive gravities. In this case the two Klein-Gordon operators appearing in (\ref{fourthorder}) are degenerate and apart from a spin-0 excitation, the spectrum also contains a logarithmic mode that obeys:
\begin{equation}
\left(\Box - m^2 \right)^2 \phi^{\mathrm{log}} = 0 \,, \qquad \left(\Box - m^2 \right) \phi^{\mathrm{log}} \neq 0 \,.
\end{equation}
In the AdS/CFT correspondence, the conformal dimension of an operator dual to a massive scalar field is related to the mass of the scalar. As in the critical limit $m_2^2 \rightarrow m_1^2 = m^2$, the mass degenerates, one expects that the operators dual to the logarithmic mode and the scalar mode with mass $m^2$ will have degenerate conformal dimension and form a logarithmic pair.

In the following, we will not work with the four-derivative formulation of the model. Instead, we will introduce an auxiliary scalar field $\phi_2$ to lower the number of derivatives from four to two. The action (for generic $m_1^2$, $m_2^2$) we will consider is given by
\begin{eqnarray} \label{actnoncrit}
S & = & \int d^{d+1}x \sqrt{g}\, \Big(-\frac{1}{2} (m_1^2 - m_2^2)
\left(\partial_\mu \phi_1 \partial^\mu \phi_1 + m_1^2
\phi_1^2\right) -
\partial_\mu \phi_1
\partial^\mu \phi_2
\nonumber \\ & & \qquad - m_1^2 \phi_1 \phi_2 - \frac{1}{2}\phi_2^2
\Big)\,.
\end{eqnarray}
Upon diagonalization this action describes two spin-0 modes with masses $m_1^2$ and $m_2^2$. The kinetic terms will have opposite signs, so the theory is always non-unitary. This is reminiscent of higher dimensional non-critical massive gravities.
Upon eliminating the auxiliary field $\phi_2$, this action leads to the equation of motion (\ref{fourthorder}). At the critical point $m_1^2 = m_2^2 = m^2$, the action reduces to \cite{Ghezelbash:1998rj,Kogan:1999bn}
\begin{equation}\label{bulkaction}
S = -\int d^{d+1}x \sqrt{g}\,\left(\partial_{\mu} \phi_1 \partial^{\mu} \phi_2 + m^2 \phi_1\phi_2 + \frac12 \phi_2^2 \right) \,.
\end{equation}
The equations of motion are then given by
\begin{eqnarray}\label{eom}
\left(\Box - m^2 \right) \phi_1 = \phi_2, \qquad \left(\Box - m^2 \right) \phi_2 = 0 \,,
\end{eqnarray}
which upon elimination of $\phi_2$  lead to a  degenerate fourth-order equation for $\phi_1$.

From now on, we will consider the bulk action (\ref{bulkaction}) and
equations of motion \eqref{eom} in the background of the
anisotropically scale invariant Lifshitz metric \eqref{Lifshitz}. We
will assume that we can ignore the backreaction of the massive
scalar on the metric. This assumption is justified when the scalar
field equations decouple from the metric equations of motion at
least asymptotically up to the order of coefficients that contribute
to the divergent terms in the bulk action \cite{Bianchi:2001kw}.

To find the non-singular bulk field configurations $\phi_i (r,t,{\bf x})$, with $i=1,2$, for any smooth boundary value $\phi_{i(0)}(t,{\bf x})$ we need to find the bulk-to-boundary propagators $G_{ij}(r,t,{\bf x};0,t',{\bf x}')$, so that:
\begin{equation}\label{bulktoboundry}
\phi_i (r,t,{\bf x}) = \sum_{j=1}^2 \int d^{d-1}{\bf x}'dt' \phi_{j(0)}(t',{\bf x}') G_{ij}(r,t, {\bf x}; 0, t',{\bf x}')\,.
\end{equation}
It is  convenient to work in Fourier space, where we transform $t$ into $\omega$ and ${\bf x}$ into ${\bf k}$. Now eq.~\eqref{bulktoboundry} reads:
\begin{equation}\label{fieldsFourier}
\phi_i (r,\omega,{\bf k}) =  \sum_{j=1}^2 \phi_{j(0)}(\omega,{\bf k}) G_{ij}(r,\omega, {\bf k})
\end{equation}
The bulk to boundary propagators $G_{ij}(r,\omega,{\bf k})$ satisfy the differential equations, for $r \ne 0$,\,:
\begin{eqnarray}
\left(\Box-m^2\right)G_{22} = 0, & \qquad & \left(\Box-m^2\right)G_{21} = 0, \label{green1}\\
\left(\Box-m^2\right)G_{11} = G_{21}, & \qquad & \left(\Box-m^2\right)G_{12} = G_{22}, \label{green2}
\end{eqnarray}
with
\begin{eqnarray}
\lefteqn{\left(\Box-m^2\right)G(r,\omega,{\bf k})} \\
 & = & r^2 \partial_r^2 G(r,\omega,{\bf k}) - (d+z-2)r \partial_r G(r,\omega,{\bf k}) - (r^{2z}\omega^2 +r^2|{\bf k}|^2 + m^2)G(r,\omega,{\bf k})\,. \nonumber
\end{eqnarray}
We impose the boundary conditions $G_{ij}(0,\omega,{\bf k})=\delta_{ij}$. Furthermore,
we have set $L=1$ for convenience. This parameter can always be  re-introduced by dimensional analysis.

We note that $\phi_1$ is the fundamental field that satisfies a
degenerate fourth-order equation of motion whereas $\phi_2$ is an
auxiliary field, needed to rewrite the equation of motion in terms
of a second-order differential equation. The most general solution
for $\phi_1$ is therefore a superposition of a mode annihilated by
acting on it with the Klein-Gordon operator once (the scalar mode)
and a mode annihilated by acting twice with the Klein-Gordon
operator (the logarithmic mode). Writing out
eq.~\eqref{fieldsFourier} for $\phi_1$ we have now two options.
Either $\phi_{1(0)}G_{11}$ is the scalar mode and
$\phi_{2(0)}G_{12}$ the logarithmic mode or vice versa. These two
options correspond to the freedom we have in coupling the sources to
the dual operators. We can either choose to couple $\phi_{1(0)}$ to
the scalar operator and $\phi_{2(0)}$ to its logarithmic partner or
vice versa. There is no difference in the physics between the two
options. We fix this ambiguity by taking $G_{11}=G_{22}=G$ and
$G_{21}=0$ so that eq.~\eqref{fieldsFourier} becomes:
\begin{align}\label{fields}
\phi_1 (r,\omega,{\bf k}) = & \; \phi_{1(0)}(\omega, {\bf k}) G(r,\omega, {\bf k}) + \phi_{2(0)}(\omega, {\bf k}) G_{12}(r,\omega, {\bf k}), \\
 \phi_{2}(r,\omega,{\bf k}) = &\; \phi_{2(0)}(\omega, {\bf k}) G(r,\omega, {\bf k}). \label{fields2}
\end{align}
Acting with one Klein-Gordon operator on $\phi_1$ will annihilate the $\phi_{1(0)}G$ term. This term therefore represents the scalar mode.
The remaining $(\Box - m^2)\phi_{2(0)}G_{12}$ term is equal to $\phi_2$ and consequently is eliminated by acting on it with a second Klein-Gordon operator. Therefore, this term represents the logarithmic mode. From the above it is  clear that $\phi_{1(0)}$ couples to a scalar operator $\mathcal{O}^s_{\Delta}$ and that $\phi_{2(0)}$ couples to its logarithmic partner $\mathcal{O}^{\rm log}_{\Delta}$ where $\Delta$ is the common conformal dimension of the two operators.

The bulk-to-boundary propagator generally has two independent
solutions. These solutions can be divided into modes which are
regular in the interior (for $r \rightarrow \infty$) and singular
modes. Since the singular modes diverge rapidly in the interior, it
is no longer safe to assume that their backreaction to the metric
can be ignored. All singular modes will therefore be discarded.

An expansion of the field near the boundary ($r \rightarrow 0$) allows us to also distinguish between the non-normalizable modes
$\phi_{i(0)}$ and the normalizable modes $\tilde{\phi}_{i(0)}$:
\begin{equation}\label{expansion}
\phi_i (r,\omega,{\bf k}) = \phi_{i(0)} (\omega,{\bf k}) r^{\Delta_-}(1 + \ldots)+ \tilde{\phi}_{i(0)} (\omega,{\bf k}) r ^{\Delta_+} (1 + \ldots)\,,
\end{equation}
where the dots indicate higher powers of $r$ within the brackets and $\Delta_+ \geq \Delta_-$ are the two roots of the quadratic equation
\begin{equation}\label{mass}
\Delta (\Delta - (d+z-1)) = m^2\,,
\end{equation}
i.e.
\begin{align}
\Delta_{\pm} & = \frac12 \left((d+z-1) \pm \sqrt{(d+z-1)^2 + 4m^2}\right)\,.
\end{align}
Note that by requiring that $\Delta_+ \geq \Delta_-$ we are assuming that  $\Delta_+ \geq (d+z-1)/2$. According to the standard AdS/CFT dictionary
 the non-normalizable mode $\phi_{i(0)}$ is the source for the dual field theory operator, while the normalizable mode $\tilde{\phi}_{i(0)}$ is related to the one-point function of the dual operator with conformal weight $\Delta=\Delta_+$.

Since the conformal dimension is related to the mass of the scalar field in the bulk, the limit where the mass of the scalar fields $\phi_1$ and $\phi_2$ degenerates corresponds to a degenerate conformal dimension for the dual operators. This is precisely what we need for a logarithmic conformal field theory, since operators with a degenerate conformal dimension will form a logarithmic pair with a non-diagonalizable Jordan cell. This degeneracy should not be confused with the degeneracy between $\Delta_+$ and $\Delta_-$ plus even integers \cite{Skenderis:2002wp} (see \cite{Taylor:2008tg} for the non-relativistic extension). We will comment briefly here on this kind of degeneracies.

The form of the power series in eq.~\eqref{expansion} can be determined by solving the equations of motion order by order in $r$. In our case, it is an expansion in  $r^{2k}$ and  $r^{2zl}$, with $k,l \in \mathbb{Z}$. Therefore, whenever $\Delta_+ - \Delta_-$ is an even integer or a multiple of $2z$, the corresponding term in the expansion of $\Delta_-$ will degenerate with the leading term in the expansion of $\Delta_+$ and a logarithmic term needs to be introduced at order $r^{\Delta_+}$. We can relate this to a value of the scalar field mass as follows:
\begin{equation}
\Delta_+ - \Delta_- = \sqrt{(d+z-1)^2 + 4m^2} = 2(k+lz), \qquad k,l \in \mathbb{Z}.
\end{equation}
The special case where $\Delta_+ = \Delta_- = (d+z-1)/2$ saturates the Breitenlohner-Freedman bound
\begin{equation} \label{BFbound}
m^2 \ge -(d+z-1)^2/4\,.
\end{equation} In
 that case the asymptotic expansion acquires a logarithmic term at leading order, because the two $\Delta$'s degenerate:
\begin{equation}\label{BFsaturated}
\phi_i (r,\omega,{\bf k}) = r^{\Delta} \left(\phi_{i(0)}
(\omega,{\bf k}) + \ldots + \log(r) \left(
\tilde{\phi}_{i(0)}(\omega,{\bf k})  +  \ldots \right) \right)\,.
\end{equation}
In the presence of this kind of degeneracies, one needs to take
additional logarithmic counterterms into account in order to get
finite correlation functions. Analogous to the discussions in
\cite{Skenderis:2002wp}, this will result in a term in the one-point
function which is a local function of the sources. At the level of
the higher-point functions these will correspond to contact terms.
For the sake of simplicity we will  restrict ourselves to those
values of $m^2$ for which no logarithmic terms arise in the
expansion of $\phi_i$ due to this kind of degeneracies. In this work
we only consider the consequences of the degeneracy of the scalar
field masses $m_1$ and $m_2$. Therefore, the results presented in
section \ref{twopoint} hold for general $m^2$ only up to contact
terms in the two-point correlation functions.

\subsection{An Example: $z=2$}\label{z=2}

To find an explicit expression for the logarithmic mode we first need to find an exact solution for the scalar mode.
Such a solution is available for the case $z=2$ \cite{Kachru:2008yh}.  We therefore consider that example in this subsection.
The solution of the homogeneous Klein-Gordon equation \eqref{green1} with $G_{11}=G_{22}=G$ and $G_{21}=0$ is given by:
\begin{equation}
G(r,\omega,{\bf k})\ \propto \ r^{\Delta} e^{-\frac{1}{2}\omega r^2 } U\left(\frac{|{\bf k}|^2+(2\Delta-(d-1))\omega}{4\omega},\Delta-\frac{d-1}{2},\omega r^2\right),
\end{equation}
where we now have that:
\begin{equation}
\Delta =  \frac12\left(d+1 + \sqrt{(d+1)^2+4m^2}\right).
\end{equation}
$U(a,b,x)$ is the confluent hypergeometric function and the constant
of proportionality can be determined by requiring that $G(\epsilon,
\omega,{\bf k}) = 1$ on the regulated boundary $r = \epsilon$. We
have found $\phi_2(r,\omega,{\bf k})$
\begin{equation}
\phi_{2}(r,\omega,{\bf k}) =  \phi_{2(0)}(\omega,{\bf k}) G(r,\omega,{\bf k}),
\end{equation}
which can be expanded near the boundary as:
\begin{align}\label{gexpand}
\phi_2 (r,\omega,{\bf k})  = & \phi_{2(0)}(\omega,{\bf k}) r^{d+1-\Delta} \left[ 1+ \ldots \right]  \\
 + \nonumber & \phi_{2(0)}(\omega, {\bf k}) \frac{\Gamma\left(\frac{d+1}{2}-\Delta\right) \Gamma \left(\frac{|{\bf k}|^2+(2\Delta-(d-1))\omega}{4\omega} \right)}{\Gamma\left(\Delta-\frac{d+1}{2}\right) \Gamma\left(\frac{|{\bf k}|^2-(2\Delta-(d+3))\omega}{4\omega} \right)}\omega^{\Delta-\frac{d+1}{2}} r^{\Delta} \left[ 1 +  \ldots \right]\,.
\end{align}
Now we still need to find $G_{12}$. For this we use a trick inspired by \cite{Kogan:1999bn}. The equation which determines $G_{12}$ is:
\begin{equation}\label{g11}
\left( \Box - m^2 \right) G_{12}(r,\omega,{\bf k}) = G(r,\omega,{\bf k})\,.
\end{equation}
From  eq.~\eqref{mass} it follows  that $\left[(\Box-m^2),d/d\Delta\right] = dm^2/d\Delta= 2\Delta-(d+1)$
 where we have used that the Lifshitz metric does not depend on the conformal dimension $\Delta$.
Using that $(\Box -m^2) G(r,\omega,{\bf k})=0$ we can therefore write $G$ as:
\begin{equation}
G = \frac{1}{2\Delta-(d+1)} \left[ \left(\Box - m^2\right), \frac{d}{d\Delta} \right]G = \frac{1}{(2\Delta-(d+1))}\left( \Box -m^2 \right) \frac{d}{d\Delta} G\,.
\end{equation}
Comparing this with eq.~\eqref{g11} we derive the following expression of $G_{12}$ in terms of the derivative of $G$ with respect to $\Delta$\,:\footnote{Note that this method is identical to the method employed in \cite{Grumiller:2008qz} to find the log modes of TMG at the critical point, albeit adjusted for scalar fields. Here one takes the limit $m_1^2 \rightarrow m_2^2$ of $(\phi_1(m_1^2)-\phi_2(m_2^2))/(m_1^2-m_2^2)$.}
\begin{equation}
G_{12}(r,\omega,{\bf k}) = \frac{1}{2\Delta-(d+1)}\frac{d}{d\Delta} G(r,\omega, {\bf k})\,.
\end{equation}
The derivative of the confluent hypergeometric function is not so easy to find. However, for our purposes, it is sufficient to derive the near boundary expansion of this derivative. The expression for this expansion can be found by taking  the derivative of the expansion \eqref{gexpand}.

According to  eq.~\eqref{fields} we have that
\begin{equation}
\phi_1 (r,\omega, {\bf k}) = \phi_{1(0)}(\omega,{\bf k}) G(r,\omega,{\bf k}) +  \phi_{2(0)}(\omega,{\bf k}) \frac{1}{2\Delta-(d+1)}\frac{d G(r,\omega,{\bf k})}{d\Delta}\,.
\end{equation}
This finally leads to the following near-boundary expansion for
$\phi_1$\,:
\begin{align}\label{gexpand2}
\phi_1 (r,\omega, {\bf k})  = & \left( \phi_{1(0)}(\omega,{\bf k}) + \phi_{2(0)}(\omega, {\bf k}) \frac{1}{((d+1)-2\Delta)} \log(r) \right) r^{d+1-\Delta} \left[ 1+ \ldots \right] \\
 + \nonumber & \bigg( \phi_{1(0)}(\omega, {\bf k}) - \phi_{2(0)}(\omega,{\bf k})\frac{1}{(d+1-2\Delta)}\Big( \log(r) + \log(\omega)  \\
 \nonumber & - \psi\left(\tfrac{d+1}{2}-\Delta\right) - \psi\left(\Delta-\tfrac{d+1}{2}\right) + \tfrac12 \psi\left(\tfrac{|{\bf k}|^2+(2\Delta-(d-1))\omega}{4\omega}\right) + \tfrac12 \psi\left(\tfrac{|{\bf k}|^2-(2\Delta-(d+3))\omega}{4\omega} \right) \Big) \bigg) \\ \nonumber
 & \times \frac{\Gamma\left(\frac{d+1}{2}-\Delta\right) \Gamma \left(\frac{|{\bf k}|^2+(2\Delta-(d-1))\omega}{4\omega}\right)}{\Gamma\left(\Delta-\frac{d+1}{2}\right) \Gamma\left(\frac{|{\bf k}|^2-(2\Delta-(d+3))\omega}{4\omega} \right)}\omega^{\Delta-\frac{d+1}{2}} r^{\Delta}  \left[ 1 +  \ldots \right]\,,
\end{align}
where $\psi(x)$ is the digamma function defined by $\psi(x) = \Gamma'(x)/\Gamma(x)$.

\section{Two point correlation functions} \label{twopoint}
Having obtained, for a specific example, the explicit expression for
the logarithmic modes we now proceed to relate these solutions to
operators on the boundary of the Lifshitz spacetime. For this we
need to apply the holographic renormalization procedure
\cite{Skenderis:2002wp}. In subsection 3.1 we first briefly review
some aspects of this procedure which will be needed later on. Next,
in subsection 3.2, we will calculate the two-point correlation
functions for an AdS background, i.e. $z=1$ and for the example
discussed in subsection 2.2, i.e. $z=2$. We will show that in both
 examples the two-point functions satisfy the defining properties of a
relativistic and non-relativistic LCFT, respectively.

\subsection{Holographic Renormalization}

For the purpose of this subsection  we may switch back to general values of $z$. Only in the next subsection
 we will specify this value. Following the AdS/CFT correspondence, we couple the boundary values of the scalar field to operators in the field theory:
\begin{equation}\label{coupling}
\int d^{d-1}xdt\;  \phi_{1(0)} \cO_{\Delta}^{\rm s} + \beta \phi_{2(0)} \cO_{\Delta}^{\rm log},
\end{equation}
where $\beta$ is a normalization parameter which we will fix later on. To precisely compute the two-point function we need to get rid of the divergences in the bulk fields as we move towards the boundary. We can do so by means of a holographic renormalization of the action \eqref{bulkaction}. Following \cite{Skenderis:2002wp} we first compute the on-shell action $S_{\rm reg}$ on a regulated surface $r=\epsilon$, using a near boundary expansion of the fields. Then we identify the divergent terms in this action as a function of the sources $\phi_{i(0)}$ and write down the counterterm action $S_{\rm ct}$ as minus these divergent terms. The counterterm action cannot be written as a covariant expression; it obeys the same anisotropic scaling as the Lifshitz background. Of course the limit $z \rightarrow 1$ should reduce to the AdS results which does allow a covariant expression. Once the counterterm action is obtained, this can be subtracted at the regulated surface to obtain the subtracted action $S_{\rm sub}$ which has by construction a finite limit for $\epsilon \rightarrow 0$.

Following the AdS/CFT dictionary, the one-point correlation functions can be obtained by functional differentiation of the on-shell action with respect to the sources:
\begin{equation}\label{onepoint}
\vev{\cO_{\Delta}^i(t,{\bf x})} =  \frac{\delta S_{\rm sub}}{\delta \phi_{i(0)}(t,{\bf x})} \bigg|_{\phi_{i(0)}=0}\,.
\end{equation}
Since the subtracted action is expressed in terms of the bulk fields $\phi_1$ and $\phi_2$ on the regulated boundary, we need to write the above
expression for the one-point correlation functions in terms of derivatives with respect to the bulk fields and afterwards take the limit $\epsilon \rightarrow 0$. To rewrite sources in
 terms of bulk fields we consider the near-boundary expansions of the bulk fields $\phi_1$ and $\phi_2$:
\begin{align} \label{phi1exp}
\phi_1  = & \left( \phi_{1(0)} + \alpha \phi_{2(0)} \log \, r \right) r^{d+z-1-\Delta} + \left( \phi_{1(2)} + \alpha \phi_{2(2)} \log \, r \right) r^{d+z+1-\Delta} \\
\nonumber & + \left( \phi_{1(2z)} + \alpha \phi_{2(2z)} \log \, r \right) r^{d+3z-1-\Delta} + \ldots + \left( \tilde{\phi}_{1(0)} - \alpha \tilde{\phi}_{2(0)} \log \, r \right) r^{\Delta} + \ldots \\
\label{phi2exp} \phi_2  = & \, \phi_{2(0)} r^{d+z-1-\Delta} +  \phi_{2(2)} r^{d+z+1-\Delta} + \phi_{2(2z)} r^{d+3z-1-\Delta} + \ldots + \tilde{\phi}_{2(0)} r^{\Delta} + \ldots\,,
\end{align}
where $\alpha$ is given by
\begin{equation}
\alpha = \frac{1}{(d+z-1-2\Delta)}\,.
\end{equation}
We can use the leading order terms in this expansion to write \eqref{onepoint} in terms of a functional derivative with respect to the bulk fields $\phi_1,\phi_2$:
\begin{align}
\vev{ \cO_{\Delta}^{\rm s}(t,{\bf x})} & = \lim_{\epsilon \rightarrow 0} \left( \frac{1}{\sqrt{\gamma}}\frac{1}{\epsilon^{\Delta}} \frac{\delta S_{\rm sub}}{\delta \phi_1(\epsilon,t,{\bf x})} \right) \\
\beta \vev{ \cO_{\Delta}^{\rm log}(t,{\bf x})} & = \lim_{\epsilon \rightarrow 0} \left( \frac{1}{\sqrt{\gamma}}\frac{1}{\epsilon^{\Delta}} \left( \frac{\delta S_{\rm sub}}{\delta \phi_2(\epsilon, t,{\bf x})} + \alpha \log \epsilon \frac{\delta S_{\rm sub}}{\delta \phi_1(\epsilon, t,{\bf x})} \right) \right)\,,
\end{align}
where $\gamma_{\alpha\beta} dx^{\alpha}dx^{\beta}= dx_adx^a/\epsilon^2 + dt^2/\epsilon^{2z}$ is the induced metric on the regulated hypersurface
and $\gamma$ is its determinant.

The two-point functions are obtained by a further differentiation of the one-point functions with respect to the sources and setting the sources to zero afterwards\,:
\begin{align}
\vev{\cO_{\Delta}^{\rm s} (t,{\bf x})\cO_{\Delta}^{\rm s}(t_2,{\bf x}_2)} & = - \frac{\delta \vev{\cO_{\Delta}^{\rm s}(t,{\bf x})}}{\delta \phi_{1(0)}(t_2,{\bf x}_2)}\bigg|_{\phi_{1(0)}=0}\,, \\
\beta \vev{\cO_{\Delta}^{\rm log}(t,{\bf x}) \cO_{\Delta}^{\rm s}(t_2,{\bf x}_2)} & = - \frac{\delta \vev{\cO_{\Delta}^{\rm s}(t,{\bf x})}}{\delta \phi_{2(0)}(t_2,{\bf x}_2)}\bigg|_{\phi_{2(0)}=0} =  - \frac{\delta \vev{\cO_{\Delta}^{\rm log}(t,{\bf x})}}{\delta \phi_{1(0)}(t_2,{\bf x}_2)} \bigg|_{\phi_{1(0)}=0}\,, \\
\beta^2 \vev{\cO_{\Delta}^{\rm log}(t,{\bf x}) \cO_{\Delta}^{\rm log}(t_2,{\bf x}_2)} & = - \frac{\delta \vev{\cO_{\Delta}^{\rm log}(t,{\bf x})}}{\delta \phi_{2(0)}(t_2,{\bf x}_2)} \bigg|_{\phi_{2(0)}=0}\,.
\end{align}

We now apply the holographic renormalization procedure to the scalar model defined by the action \eqref{bulkaction}.
A partial integration of this action on a regulated surface $r= \epsilon$ near the boundary and requiring the equations of motion to hold leads to the following regularized on-shell action:
\begin{equation}
S_{\rm reg} = - \frac12 \int_{r=\epsilon} d^dx \sqrt{\gamma} \left(\phi_1 \vec{n} \cdot \vec{\nabla} \phi_2 + \phi_2 \vec{n} \cdot \vec{\nabla} \phi_1 \right)\,,
\end{equation}
where $\vec{n}$ is the vector normal to the regulated hypersurface $\vec{n}\cdot \vec{\nabla} = r \partial_r |_{r=\epsilon}$.

Without explicitly going through all the steps of the holographic renormalization procedure, we note that after a lengthy calculation
we find that the counterterm action needed to make the action \eqref{bulkaction} finite is given by:
\begin{align}\label{counterterms}
S_{\rm ct} = &  \int_{r=\epsilon} d^dx \sqrt{\gamma}\bigg( (d+z-1-\Delta)\phi_1 \phi_2 + \frac12 \alpha \phi_2 \phi_2 \\
\nonumber & - a_2 \left(\frac12 \left(\phi_1 \partial^a\partial_a \phi_2 + \phi_2 \partial^a\partial_a \phi_1\right) - a_2 \alpha \phi_2\partial^a \partial_a \phi_2\right) \\
\nonumber &  - a_{2z} \left(\frac12 \left(\phi_1 \partial^t\partial_t \phi_2 + \phi_2 \partial^t\partial_t \phi_1\right) - a_{2z} \alpha \phi_2\partial^t \partial_t \phi_2\right) + \cO \! \left( \phi_i \partial_a^4 \phi_i \right) \bigg)\,,
\end{align}
with $a=1,\cdots ,d-1$ and $a_2$ and $a_{2z}$ given by
\begin{equation}
 a_2 =  \frac{1}{(d+z+1-2\Delta)}, \hspace{1cm} a_{2z} = \frac{1}{(d+3z-1-2\Delta)}\,.
\end{equation}
We note that all indices in the derivatives are raised and lowered with the induced metric on the boundary $\gamma_{\alpha\beta}$.

In \eqref{counterterms} we took terms up to order $\cO \left( \phi_i \partial_a^4 \phi_i \right)$ into account. In the near-boundary expansions \eqref{phi1exp}, \eqref{phi2exp} the normalizable modes are of order $\epsilon^{\Delta}$, so all the terms with a lower power than $\epsilon^{\Delta}$ are going to contribute to the counterterm action. The precise number of counterterms we need to add depends on the value of $\Delta$. This value of
$\Delta$ is restricted as follows:
\begin{equation}
\frac12(d+z-1) \leq \Delta \leq  d+z-1\,.
\end{equation}
The upper limit follows from the observation that if $\Delta \geq
d+z-1$ then the operator is irrelevant and, according to
\cite{vanRees:2011fr},  it is no longer safe to ignore the
backreaction of  the scalar sector on the gravitational background.
The lower limit follows from the Breitenlohner-Freedman bound
\eqref{BFbound}. For our purposes, taking counterterms into account
up to order $\epsilon^{d+z-1}$ is sufficient. In
eq.~\eqref{counterterms} we have only written down the first couple
of terms. These are sufficient for $d=2$ and $z=1,2$. These terms
illustrate that the counterterm action cannot be written down
covariantly, but instead respects the anisotropic scale invariance.
For larger values of $d$ and $z$ we need to take more counterterms
into account, but the analysis is similar and can be extended
straightforwardly. The renormalized one-point correlation functions
do not change as long as the degeneracy discussed at the end of
section \ref{genfeat} is absent. Their expressions are given by:
\begin{align}
\langle \cO_{\Delta}^{\rm s} (t,{\bf x}) \rangle = & \, (d+z-1 - 2\Delta) \tilde{\phi}_{2(0)}(t,{\bf x})\,, \\
\beta \langle \cO_{\Delta}^{\rm log} (t,{\bf x}) \rangle = & \,
(d+z-1 - 2\Delta) \tilde{\phi}_{1(0)}(t,{\bf x})\,,
\end{align}
where $\beta$ is the normalization parameter that appeared in
\eqref{coupling}.

\subsection{Two-point Correlation Functions}

The two-point functions can now be obtained from the exact solutions
to the field equations. Once the exact solution is found, we can
expand it near the boundary and find the expressions for
$\tilde{\phi}_{i(0)}$ linearly in the sources  $\phi_{i(0)}$. To
find exact solutions  we need to specify the value of $z$. Below we
discuss two examples.

\subsubsection{Example 1: $z=1$}
 We first consider $z=1$, i.e.~the $d$ dimensional LCFT dual to $d+1$ dimensional Anti-de Sitter. The solution to the homogeneous Klein-Gordon equation which is regular everywhere in the interior in Fourier space is:
 \begin{equation}
 G(r,k) \propto r^{\frac{d}{2}} K_{\frac{1}{2} \sqrt{d^2+4 m^2}}(|k|
 r)\,,
 \end{equation}
 where  $k = \{\omega,{\bf k} \}$  is now a $d$ component vector with length $|k|$ and $K_n(z)$ is the modified Bessel function of the second kind. The constant of proportionality is determined by taking $G(\epsilon,k) = 1$ on the regulated boundary.

Repeating the steps outlined in section \ref{z=2} and applying the holographic renormalization outlined above for $z=1$ we find that the correlation functions expressed in Fourier space are:
\begin{align}
\langle \cO_{\Delta}^{\rm s} (k) \cO_{\Delta}^{\rm s}(-k) \rangle &  = 0\,,  \label{2ptads1} \\
 \beta \langle \cO_{\Delta}^{\rm s}(k) \cO_{\Delta}^{\rm log}(-k) \rangle & = (2\Delta-d) |k|^{2\Delta-d} \frac{2^{d-2\Delta}\Gamma\left(\frac{d}{2}-\Delta\right)}{\Gamma\left(\Delta-\frac{d}{2}\right)}\,, \label{2ptads2} \\
\label{2ptads3} \beta^2 \langle \cO_{\Delta}^{\rm log}(k) \cO_{\Delta}^{\rm log}(-k) \rangle & =  |k|^{2\Delta-d} \frac{2^{d-2\Delta}\Gamma\left(\frac{d}{2}-\Delta\right)}{\Gamma\left(\Delta-\frac{d}{2}\right)}\big(2\log|k| \\
& \;\;\; - \log 4 - \psi (\Delta -\tfrac{d}{2}) - \psi(\tfrac{d}{2}-\Delta) \big)\,. \nonumber
\end{align}
As expected, this is precisely the structure of a relativistic LCFT \cite{Kogan:1999bn}.

\subsubsection{Example 2: $z=2$}\label{z2}
For the example worked out in subsection 2.2 with $z=2$ we can read off $\tilde{\phi}_{1(0)}$ and $\tilde{\phi}_{2(0)}$ by comparing \eqref{phi1exp} with \eqref{gexpand} and \eqref{phi2exp} with \eqref{gexpand2}. This leads to the following expressions\,:
\begin{align}
\tilde{\phi}_{1(0)} & = \omega^{\Delta-\frac{d+1}{2}} \frac{\Gamma\left(\frac{d+1}{2}-\Delta\right) \Gamma \left(\frac{|{\bf k}|^2+(2\Delta-(d-1))\omega}{4\omega}\right)}{\Gamma\left(\Delta-\frac{d+1}{2}\right) \Gamma\left(\frac{|{\bf k}|^2-(2\Delta-(d+3))\omega}{4\omega} \right)} \bigg(\phi_{1(0)} -  \phi_{2(0)}  \frac{1}{d+1-2\Delta}  \Big( \log (\omega)   \\
\nonumber & - \psi\left(\tfrac{d+1}{2}-\Delta\right) - \psi\left(\Delta-\tfrac{d+1}{2}\right) + \tfrac12 \psi\left(\tfrac{|{\bf k}|^2+(2\Delta-(d-1))\omega}{4\omega}\right) + \tfrac12 \psi\left(\tfrac{|{\bf k}|^2-(2\Delta-(d+3))\omega}{4\omega} \right) \Big) \bigg)\,, \\
\tilde{\phi}_{2(0)} & = \phi_{2(0)} \frac{1}{d+1-2\Delta} \omega^{\Delta-\frac{d+1}{2}} \frac{\Gamma\left(\frac{d+1}{2}-\Delta\right) \Gamma \left(\frac{|{\bf k}|^2+(2\Delta-(d-1))\omega}{4\omega}\right)}{\Gamma\left(\Delta-\frac{d+1}{2}\right) \Gamma\left(\frac{|{\bf k}|^2-(2\Delta-(d+3))\omega}{4\omega} \right)}\,.
\end{align}
Following the general procedure outlined in the previous subsection we find that the two-point functions are:
\begin{align}
\langle \cO_{\Delta}^{\rm s} (\omega, {\bf k}) \cO_{\Delta}^{\rm s}(-\omega,-{\bf k}) \rangle & = 0  \label{2pt1}\,, \\
 \beta \langle \cO_{\Delta}^{\rm s}(\omega,{\bf k}) \cO_{\Delta}^{\rm log}(-\omega,- {\bf k}) \rangle & =  (2\Delta- (d+1)) \omega^{\Delta-\frac{d+1}{2}} \frac{\Gamma\left(\frac{d+1}{2}-\Delta\right) \Gamma \left(\frac{|{\bf k}|^2+(2\Delta-(d-1))\omega}{4\omega}\right)}{\Gamma\left(\Delta-\frac{d+1}{2}\right) \Gamma\left(\frac{|{\bf k}|^2-(2\Delta-(d+3))\omega}{4\omega} \right)} \label{2pt2}\,, \\
 \label{2pt3} \beta^2 \langle \cO_{\Delta}^{\rm log}(\omega,{\bf k}) \cO_{\Delta}^{\rm log}(-\omega,- {\bf k}) \rangle
& =   \omega^{\Delta-\frac{d+1}{2}} \frac{\Gamma\left(\frac{d+1}{2}-\Delta\right) \Gamma \left(\frac{|{\bf k}|^2 + (2\Delta-(d-1))\omega}{4\omega} \right)}{\Gamma\left(\Delta-\frac{d+1}{2}\right) \Gamma\left(\frac{|{\bf k}|^2-(2\Delta-(d+3))\omega}{4\omega} \right)}  \bigg( \log \omega
\end{align}
\begin{equation}
 \nonumber  - \psi\left(\tfrac{d+1}{2}-\Delta\right) - \psi\left(\Delta-\tfrac{d+1}{2}\right) + \tfrac12 \psi\left(\tfrac{|{\bf k}|^2+(2\Delta-(d-1))\omega}{4\omega}\right) + \tfrac12 \psi\left(\tfrac{|{\bf k}|^2-(2\Delta-(d+3))\omega}{4\omega} \right) \bigg) \,.
\end{equation}
The correlation function \eqref{2pt2} agrees with the two point function for a massive scalar field in a Lifshitz background found in \cite{Kachru:2008yh} and later by means of holographic renormalization in \cite{Taylor:2008tg}.

\subsection{Comparison with LCFT's}
In the relativistic case, a general logarithmic conformal field
theory of rank 2 (i.e.~only one logarithmic partner) has two-point
correlation functions which are restricted by the conformal symmetry
to be \cite{Gurarie:1993xq}:
\begin{align}
\vev{\cO^{\rm s}(x) \cO^{\rm s}(y) } & = 0\,, \label{lcft2pt1} \\
\vev{\cO^{\rm log}(x) \cO^{\rm s}(y)} & = \frac{c}{|x-y|^{2\Delta}}\,, \label{lcft2pt2} \\
\vev{\cO^{\rm log}(x) \cO^{\rm log}(y)} & = \frac{1}{|x-y|^{2\Delta}}\left(-2c \log |x-y| + \lambda \right)\,, \label{lcft2pt3}
\end{align}
where the constant $c$ is determined by the normalization of $\cO^{\rm log}$ and the constant $\lambda$ can be changed by the rescaling $ \cO^{\rm log} \rightarrow \cO^{\rm log} + \cO^{\rm s}$.

To re-write these expressions in Fourier space we use the fact that the Fourier transform of a power law in $d$ dimensions is given by another power law. Explicitly, one finds:
\begin{align}
\vev{ \cO^{\rm log} (k) \cO^{\rm s}(-k)} &  = \frac{1}{(2\pi)^{d/2}} \int d^d x\; e^{-ik\cdot x} \frac{c}{|x|^{2\Delta}}\nonumber \\
& = 2^{d/2-2\Delta } \frac{\Gamma \left(\frac{d}{2} - \Delta\right)}{\Gamma\left(\frac{d}{2}\right) \Gamma\left(1+\Delta-\frac{d}{2}\right) } c |k|^{2\Delta-d}
\end{align}
and:
\begin{align}
\vev{ \cO^{\rm log} (k) \cO^{\rm log}(-k)} &  = \frac{1}{(2\pi)^{d/2}} \int d^d x\; e^{-ik\cdot x} \frac{c}{|x|^{2\Delta}}\left(- 2 \log x + \lambda \right) \nonumber  \\
& = 2^{d/2-2\Delta-1} \frac{\Gamma \left(\frac{d}{2} - \Delta \right)}{\Gamma\left(\frac{d}{2}\right) \Gamma\left(1+\Delta-\frac{d}{2} \right) } c |k|^{2\Delta-d} \bigg( 2 \log |k| \nonumber \\
 & \;\;\; - \log 4 - \psi \left(\frac{d}{2} - \Delta \right) - \psi \left(1+\Delta-\frac{d}{2} \right) + 2 \lambda \bigg)\,.
\end{align}
If we compare these expressions with the correlation functions obtained in section 3.2.1 from the holographic calculation with bulk AdS space we find that they agree and the standard normalization is obtained by choosing $\beta = 1/(\Delta - d/2)$.

For non-relativistic field theories the two-point functions are less restricted by the symmetry group. Invariance under time and space translations and spatial rotations restrict the two point correlation functions to be functions of only $|t-t'|$ and $|{\bf x}-{\bf x}'|$. The non-relativistic scale transformations then further restrict the general two-point function of two operators with scaling dimensions $\Delta_1$ and $\Delta_2$ to be:
\begin{equation}
\vev{ \cO_{\Delta_1}(t_1,{\bf x}_1) \cO_{\Delta_2}(t_2,{\bf x}_2)} =
\frac{1}{|{\bf x}_1-{\bf x}_2|^{\Delta_1 + \Delta_2}} f(\chi) =
\frac{1}{|t_1-t_2|^{(\Delta_1+\Delta_2)/z}} f'(\chi)\,,
\end{equation}
where $f(\chi),f'(\chi)$ are arbitrary functions of the scale invariant variable $\chi = \frac{|{\bf x}_1-{\bf x}_2|^z}{|t_1-t_2|}$.

If we compare this with the correlation functions found in section \ref{z2} we see that they show the appropriate scaling behavior. By analogy to the AdS case, this suggests that the general structure of the non-relativistic LCFT is:
\begin{align}
\vev{\cO^{\rm s}(t_1,{\bf x}_1) \cO^{\rm s}(t_2,{\bf x}_2) } & = 0\,, \\
\vev{\cO^{\rm log}(t_1,{\bf x}_1) \cO^{\rm s}(t_2,{\bf x}_2)} & = \frac{1}{|{\bf x}_1-{\bf x}_2|^{2\Delta}} f(\chi) \,,  \\
\vev{\cO^{\rm log}(t_1,{\bf x}_1) \cO^{\rm log}(t_2,{\bf x}_2)} & =
\frac{1}{|{\bf x}_1-{\bf x}_2|^{2\Delta}}\left(- g(\chi) \log |{\bf
x}_1-{\bf x}_2| + \lambda \right)\,,
\end{align}
with $\lambda$ a constant which can be changed by transforming $\cO^{\rm log} \rightarrow \cO^{\rm log} + \cO^{\rm s}$ and $f(\chi),g(\chi)$ are arbitrary functions of the scale invariant variable $\chi$.

\section{Conclusions}

In this work we considered a fourth-order derivative scalar field
configuration. Upon using an auxiliary scalar field, the model
describes two ordinary Klein-Gordon scalar fields with mass squared
$m_1^2$ and $m_2^2$ and with opposite signs of their kinetic terms.
Like in theories of massive gravity, there exists a critical case
where $m_1^2 = m_2^2 = m^2$, that exhibits a logarithmic mode, apart
from an ordinary scalar mode. In the relativistic case, when
considering a fixed AdS background, the model was shown to be dual
to a logarithmic CFT \cite{Kogan:1999bn}. Instead of considering a
fixed AdS background, in this letter we considered a
non-relativistic Lifshitz background. Just as the usual AdS/CFT
correspondence is then extended to a non-relativistic version,
likewise we suggest that the fourth-order derivative scalar model is
dual to a non-relativistic version of a logarithmic CFT. We then
employed non-relativistic holographic methods to calculate the
two-point functions of the operators sourced by the boundary value
of the scalar and the logarithmic mode. Holographic reasoning allows
one to view these correlation functions on the boundary as defining
a non-relativistic extension of a logarithmic CFT.

Although the model we discussed here involves only spin-0 degrees of
freedom, it bears a lot of resemblance with massive gravity
theories. Away from critical points, the latter describe both
massive and massless (or pure gauge for $d=3$) spin-2 degrees of
freedom. At a critical point, the massive gravitons become massless
and are replaced by logarithmic modes. At such a critical point, the
theories are conjectured to be dual to logarithmic CFTs. In view of
this similarity to critical gravity, it would be interesting to
consider critical gravities around a non-relativistic background and
obtain non-relativistic versions of the log CFTs dual to massive
gravity theories. In these log CFTs, typically the stress energy
tensor would acquire a logarithmic partner. In this respect it is of
interest to note that massive gravity theories, like Topologically
Massive Gravity and New Massive Gravity, generically exhibit
Lifshitz vacua.

Finally, it would be interesting to see whether these non-relativistic
log CFTs, obtained via holographic reasoning can also be understood
as deformations of relativistic log CFTs as it can be done for ordinary
non-relativistic CFTs \cite{Guica:2010sw}.

\section*{Acknowledgements}
S.d.H, W.M. and J.R. are financed by the Dutch stichting voor Fundamenteel Onderzoek der Materie (FOM).

\providecommand{\href}[2]{#2}\begingroup\raggedright\endgroup

\end{document}